\documentclass[twocolumn,aps,prc,superscriptaddress,showpacs,floatfix]{revtex4}
\usepackage{graphicx}

\begin{document}
\title{Jet conversions in a quark-gluon plasma}

\author{W. Liu}
\affiliation{Cyclotron Institute and Physics Department, Texas A$\&$M
University, College Station, Texas 77843-3366}
\author{C. M. Ko}
\affiliation{Cyclotron Institute and Physics Department, Texas A$\&$M
University, College Station, Texas 77843-3366}
\author{B. W. Zhang\footnote{On leave from
Institute of Particle Physics, Huazhong Normal University, Wuhan 430079,
China}}
\affiliation{Cyclotron Institute and Physics Department, Texas A$\&$M
University, College Station, Texas 77843-3366}

\begin{abstract}
Quark and gluon jets traversing through a quark-gluon plasma not
only lose their energies but also can undergo flavor conversions.
The conversion rates via the elastic $q(\bar q)g\to gq(\bar q)$ and
the inelastic $q\bar q\leftrightarrow gg$ scatterings are evaluated
in the lowest order in QCD. Including both jet energy loss and
conversions in the expanding quark-gluon plasma produced in
relativistic heavy ion collisions, we have found a net conversion of
quark to gluon jets. This reduces the difference between the nuclear
modification factors for quark and gluon jets in central heavy ion
collisions and thus enhances the $p/\pi^+$ and ${\bar p}/\pi^-$
ratios at high transverse momentum. However, a much larger net quark
to gluon jet conversion rate than the one given by the lowest-order
QCD is needed to account for the observed similar ratios in central
Au+Au and p+p collisions at same energy. Implications of our results
are discussed.
\end{abstract}


\maketitle

\section{introduction}

One of the most interesting observations in central heavy ion
collisions at the Relativistic Heavy Ion Collider (RHIC) is the
suppressed production of hadrons with large transverse momentum
\cite{adcox,adler1}. This phenomenon has been attributed to the
radiative energy loss of quark and gluon jets, produced from initial
hard scattering of incoming nucleons, as they traverse through
produced quark-gluon plasma (QGP) \cite{wang,gyulassy,wiedemann}.
Recent studies have shown that elastic scattering of quark and gluon
jets in the QGP also leads to an appreciable loss of their energies
\cite{mustafa,djordjevic1}. Because of its larger color charge, a
gluon jet is expected to lose more energy than quark and antiquark
jets.  Since the ratio of high momentum protons and antiprotons to
pions produced from the fragmentation of a gluon jet is much larger
than that from a quark jet and there are more gluon than quark jets
in proton-proton collisions, a larger gluon than quark energy loss
would lead to smaller $p/\pi^+$ and ${\bar p}/\pi^-$ ratios at high
transverse momentum in central heavy ion collisions than in
proton-proton collisions at same energy \cite{wang1}. This is in
contrast to $p/\pi^+$ and ${\bar p}/\pi^-$ ratios at intermediate
transverse momentum where they are enhanced in central heavy ion
collisions as a result of quark coalescence or recombination
\cite{hwa,greco,fries}. Experimentally, data from the STAR
collaboration have indicated, however, that $p/\pi^+$ and ${\bar
p}/\pi^-$ ratios at high transverse momentum in central Au+Au
collisions \cite{adams} approach those in p+p and d+Au collisions
\cite{adams1}. This observation implies that the ratio of final
quark and gluon jets at high transverse momentum is similar to that
of initial ones. A possible mechanism for reducing the effect due to
different quark and gluon jet energy losses in QGP is to allow
conversions between quark and gluon jets via both elastic $gq(\bar
q)\to q(\bar q)g$ and inelastic $q\bar q\leftrightarrow gg$
scatterings with thermal quarks and gluons in the QGP. The idea of a
quark jet converting to a gluon jet has previously been considered
in Refs.\cite{wang2} for deeply inelastic scattering. In the present
paper, we study this effect in the lowest-order in QCD. We find that
conversions between quark and gluon jets indeed lead to an increase
in the final number of gluon jets in central heavy ion collisions
than the case without conversions but the effect is not large enough
to explain the experimental observation. To bring the resulting
$p/\pi^+$ and ${\bar p}/\pi^-$ ratios at high transverse momentum to
the values observed in p+p collisions at same energy requires a
factor of four or larger in the increase of the net quark to gluon
jet conversion rate obtained from the lowest-order QCD.

\section{jet conversion rates in quark-gluon plasma}

The conversion rate of a quark jet to a gluon jet or vice versa due
to two-body scattering with quarks and gluons in a QGP is given by
the collisional width
\begin{eqnarray}\label{gamma}
\Gamma_C&=&\frac{1}{2E_1}\int\frac{g_2d^3{\bf p}_2}{(2\pi)^32E_2}
\frac{d^3{\bf p}_3}{(2\pi)^32E_3}\frac{d^3{\bf p}_4}{(2\pi)^32E_4}
\nonumber\\
&\times&f({\bf p}_2)[1\pm f({\bf p}_3)][1\pm f({\bf
p}_4)]\overline{|{\cal M}_{12\to 34}|^2}
\nonumber\\
&\times&(2\pi)^4\delta^{(4)}(p_1+p_2-p_3-p_4)
\end{eqnarray}
divided by $\hbar$. In the above, $E_1$ and $E_3$ are, respectively,
energies of the jet before and after conversion, while $E_2$ and
$E_4$ are those of quarks and gluons with degeneracy $g_2$ in the
QGP; $f({\bf p})$ is the thermal distribution with $+$ and $-$
referring to gluons and quarks, respectively; and $\overline{|{\cal
M}_{12\to 34}|^2}$ is the squared scattering amplitude after
averaging over the spins and colors of initial partons and summing
over those of final partons and is well-known for both the elastic
scattering $q(\bar q)g\to gq(\bar q)$ and the inelastic scattering
$q\bar q\leftrightarrow gg$. To ensure that the quark (gluon) jet is
converted to a gluon (quark) jet in elastic scattering, the gluon
(quark) in the final state is required to have a larger momentum.
Since both elastic and inelastic scattering are dominated by
forward-peaked $t-$ and backward-peaked $u-$ channel diagrams, the
momentum of the converted jet is very close to that of the initial
jet. Neglecting the effect due to the soft parton in a conversion
process is thus justified.

\begin{figure}[ht]\label{width1}
\includegraphics[width=2.2in,height=2.8in,angle=-90]{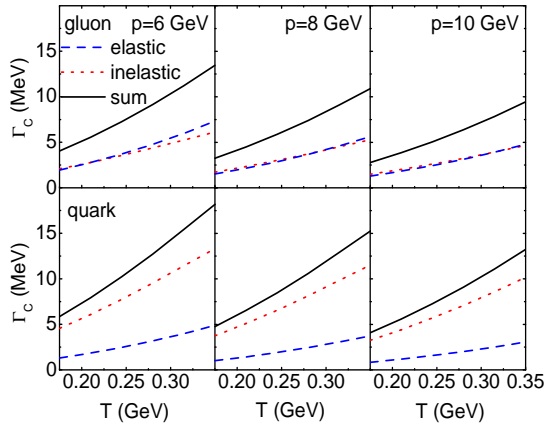}
\caption{(Color online) Collisional widths for gluon to quark jet
(upper panels) and quark to gluon jet (lower panels) conversions in QGP
due to elastic scattering $q(\bar q)g \to gq(\bar q)$ (dashed lines) and
inelastic scattering $q\bar q\leftrightarrow gg$ (dotted lines) as well as
their sum (solid lines) as functions of temperature for different quark
and gluon momenta of 6 (left panels), 8 (middle panels), and 10
(right panels) GeV/$c$.}
\end{figure}

To take into account medium effects, we include the thermal masses
$m_q=m_g/\sqrt{3}=gT/\sqrt{6}$ \cite{blazoit} for quarks and gluons
in a QGP at temperature $T$, where $g$ is the QCD coupling constant.
With $\alpha_s=g^2/4\pi=0.3$, appropriate for energy scales
considered here, calculated collisional widths for gluon to quark
jet (upper panels) and quark to gluon jet (lower panels) conversions
in a chemically equilibrated QGP are shown in Fig.1 for jets of
momentum 6 (left panels), 8 (middle panels), and 10 (right panels)
GeV/$c$. Because of larger (about a factor of two) quark than gluon
densities in the chemically equilibrated QGP with thermal quark and
gluon masses, contributions from elastic (dashed lines) and
inelastic (dotted lines) scattering to conversion of gluon jets to
quark jets are comparable, while inelastic scattering is more
important than elastic scattering for quark to gluon jet conversion.
Adding both contributions leads to a larger total conversion rate
for the quark jet than for the gluon jet, particularly at high
transverse momentum, as shown by solid lines in Fig.1.

\section{drag coefficients for jets in quark-gluon plasma}

As in our study of heavy quark energy loss \cite{weiliu}, the
momentum degradation of a quark or gluon jet in the QGP depends on
its drag coefficient, which is given by averages similar to that for
the collisional width, i.e.,
\begin{equation}
\gamma(|{\bf p}|,T)=\sum_i\langle\overline{|{\cal M}_i|^2}\rangle
-\sum_i\langle\overline{|{\cal M}_i|^2}{\bf p}\cdot{\bf p^\prime}
\rangle/|\bf p|^2,
\end{equation}
where ${\bf p}$ and ${\bf p}^\prime$ are the momenta of the jet
before and after a collision, respectively, and the sum is over all
scattering processes. Since the present study is mainly concerned
with conversions between gluon and quark jets in the QGP, we only
consider explicitly the contribution of two-body scattering to their
energy losses and mimic the effect of the more important radiative
energy loss by introducing a phenomenological multiplication factor
$K_E$.

\begin{figure}[ht]
\includegraphics[width=2.2in,height=2.2in,angle=-90]{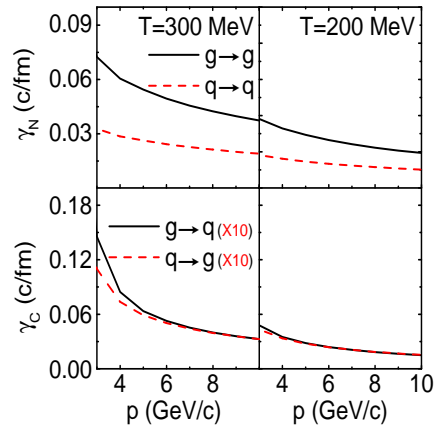}
\caption{(Color online) Drag coefficients for gluon and quark jets
in QGP due to non-conversion (upper panels) and conversion (lower
panels) two-body scattering as functions of their momentum in QGP at
temperature $T=300$ MeV (left panels) or $T=200$ MeV (right
panels).} \label{drag}
\end{figure}

In Fig.\ref{drag}, we show the gluon and quark drag coefficients
$\gamma_N$ due to non-conversion (upper panels) and $\gamma_C$ due
to conversion (lower panels) scatterings as functions of their
momentum in a QGP at temperatures $T=300$ MeV (left panels) or
$T=200$ MeV (right panels). It is seen that the drag coefficients
due to non-conversion scattering are much larger than those due to
conversion scattering. While the drag coefficient for the gluon jet
is about a factor of two large than that for the quark jet in
non-conversion scattering, they are similar in conversion
scattering. Enhancing the calculated drag coefficients due to
two-body scattering by a factor $K_E=4$, we find that both their
values and transverse momentum dependence are consistent with those
extracted from the energy loss formula derived in
Ref.\cite{gyulassy} for radiative energy losses of quark and gluon
jets in a quark-gluon plasma. We have also checked that with such
enhanced drag coefficients, the observed quenching of high
transverse momentum pions at RHIC can indeed be described.

\section{initial jet distributions and heavy ion collision dynamics}

\begin{table}[th]
\caption{Parameters for initial minijet parton distributions given
in Eq.(\ref{jet}) at midrapidity from Au+Au at $\sqrt{s_{NN}}=200$
GeV.}
\medskip
\begin{tabular}{cccc}\hline\hline
\hfil & $\quad A[({\rm GeV}/c)^{-2}]\quad$ & $\quad B[{\rm
GeV}/c]\quad$ & $\quad \alpha\quad$ \\ \hline
 {\it g} & 1440 & 1.5 & 8.0 \\ \hline
 {\it q} & 670  & 1.6 & 7.9 \\ \hline
$\bar q$ & 190  & 1.9 & 8.9 \\ \hline
\end{tabular}
\label{parameters}
\end{table}

To see the effect of conversions between quark and gluon jets on
their energy losses in QGP, we consider central Au+Au collisions at
center-of-mass energy $\sqrt{s_{NN}}=200$ GeV. The initial
transverse momentum spectra of quark, anti-quark, and gluon jets at
mid-rapidity are obtained from multiplying those from PYTHIA for p+p
collisions at same energy by the number of binary collisions ($\sim
960$) in central Au+Au collisions. The resulting transverse momentum
spectra of initial partonic jets can be parameterized by
\begin{equation}\label{jet}
\frac{dN_i}{d^2 p_T}\approx A_i/(1+p_T/B_i)^{\alpha_i}
\end{equation}
with $i=q, \bar q$, and $g$. With transverse momentum $p_T$ in unit
of GeV/$c$, values of the parameters $A_i$, $B_i$, and $\alpha_i$ in
above equation are given in Table~\ref{parameters}.

For the dynamics of formed QGP, we follow that of Ref.\cite{chen} by
assuming that it evolves boost invariantly in the longitudinal
direction but with an accelerated transverse expansion.
Specifically, its volume expands in the proper time $\tau$ according
to $V(\tau)=\pi R(\tau)^2\tau c$, where
$R(\tau)=R_0+a(\tau-\tau_0)^2/2$ is the transverse radius with an
initial value $R_0$=7 fm, $\tau_0$=0.6 fm/$c$ is the QGP formation
time, and $a=0.1c^2$/fm is the transverse acceleration. Starting
with an initial temperature $T_i=350$ MeV, time dependence of the
temperature is obtained from entropy conservation, leading to the
critical temperature $T_c=175$ MeV at proper time $\tau_c=5$ fm/$c$.

For a quark or gluon jet moving through the QGP, the rate for the
change of its mean transverse momentum $\langle p_T\rangle$ is given
by
\begin{equation}
\frac{d\langle p_T\rangle}{d\tau}=-\langle\gamma(p_T,T)p_T\rangle
\approx-\gamma(\langle p_T\rangle,T)\langle p_T\rangle.
\end{equation}
In obtaining the last expression of above equation, we have
neglected the dispersion of the jet momentum during its propagation
and thus assumed that $\langle p_T^2\rangle\approx\langle
p_T\rangle^2$.  This is expected to be a reasonable approximation
for high transverse momentum particles.

Because of conversion scatterings, a quark or gluon jet can be
converted to a gluon or quark jet with a rate given by corresponding
collisional width. These effects are modeled by introducing a large
number of test quark and gluon jets that are distributed in the
transverse plane according to that of underlying binary
nucleon-nucleon collisions. Their transverse momentum distribution
is taken to be uniform with directions pointed isotropically in the
transverse plane. A test jet of a given kind with certain transverse
momentum is assigned a probability that is proportional to the
corresponding jet momentum spectrum with the proportional constant
determined by requiring that the sum of the probabilities for all
test jets of this kind is equal to their total number. Motions of
the jets are then followed via straight trajectories. After a small
time step $\Delta\tau$, which is usually taken to be 0.1 fm/$c$, a
random number $x$ is chosen and compared with
$y=\Gamma_C\Delta\tau/\hbar$. If $x$ is smaller than $y$, then the
jet is converted to a different kind with a reduction of its
momentum $p_T$ by $\gamma_C\Delta\tau p_T$ but without changing its
associated probability. Otherwise, the jet loses a momentum
$\gamma_N\Delta\tau p_T$. This sequence of processes is repeated
until the smaller of the time when the QGP phase ends and the time
for the jet to escape the expanding QGP. The final momentum
distributions of quark and gluon jets are obtained from those of
test jets by multiplying with their associated probabilities.

\section{results}

\begin{figure}[ht]
\includegraphics[width=2.2in,height=2.2in,angle=-90]{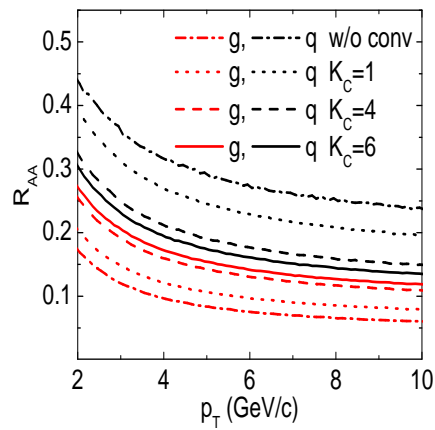}
\caption{(Color online) Nuclear modification factors for quark
(upper lines) and gluon (lower lines) jets in central Au+Au
collisions at $\sqrt{s_{NN}}=200$ GeV as functions of momentum
without (dash-dotted lines) or with different enhancement factors
$K_C=1$ (dotted lines), $K_C=4$ (dashed lines), and $K_C=6$ (solid
lines) for conversion scattering.} \label{raa}
\end{figure}

The ratio of final quark or gluon jet spectrum to its initial
spectrum, defined as its nuclear modification factor $R_{AA}$, is
shown in Fig.\ref{raa}. Upper and lower dash-dotted lines are those
for the quark and gluon jets using drag coefficients from two-body
elastic scattering that are enhanced by the factor $K_E=4$ but
without jet conversions. The smaller $R_{AA}$ for gluons than for
quarks is a result of their larger energy loss. Although final high
transverse momentum quark and gluon jets are initially produced near
the surface of the QGP fireball, quark jets tend to originate from a
region that is deeper inside the fireball than that for gluon jets
as a result of their smaller drag coefficients. With the larger
quark to gluon conversion width than that for gluon to quark
conversion, it is thus more likely for quark jets to convert to
gluon jets than for gluon jets to convert to quark jets. Including
conversions between quark and gluon jets through conversion
scatterings thus reduces the difference between the quark and gluon
$R_{AA}$ as shown by dotted lines in Fig.\ref{raa}.

\begin{figure}[ht]
\vspace{0.5cm}
\includegraphics[width=2.2in,height=3in,angle=-90]{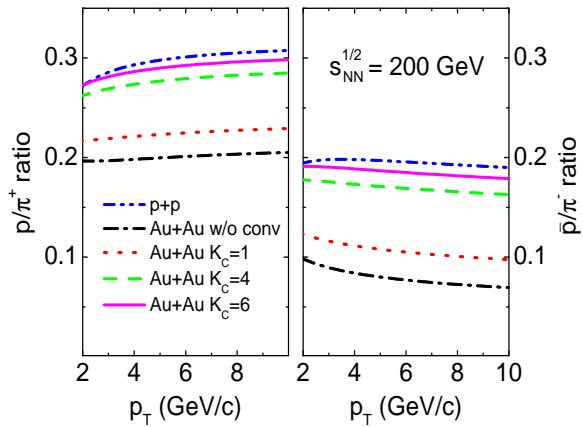}
\caption{(Color online) $p/\pi^+$ (left panel) and ${\bar p}/\pi^-$
(right panel) ratios from quark and gluon jet fragmentation in
central Au+Au collisions at $\sqrt{s_{NN}}=200$ GeV as functions of
momentum without (dash-dotted lines) or with different enhancement
factors $K_C=1$ (dotted lines), $K_C=4$ (dashed lines), and $K_C=6$
(solid lines) for conversion scattering. Dash-dot-dotted lines
correspond to p+p collisions at same energy.} \label{ratio}
\end{figure}

To obtain the proton, antiproton and pion spectra at high transverse
momentum from those of quark and gluon jets, we use the AKK
fragmentation function \cite{akk}, which has been shown to reproduce
measured charged pion, proton, and antiproton spectra at high
transverse momentum in p+p and d+Au collisions at
$\sqrt{s}_{NN}=200$ GeV. In particular, the resulting ${\bar p}/p$
ratio at high transverse momentum in p+p collisions is about 0.7 and
agrees with the observed value. While protons and antiprotons are
equally produced from gluon fragmentation, fragmentation of quark
and antiquark jets is known to produce mainly protons and
antiprotons, respectively \cite{adams,adams1,straub}. We therefore
assume that no antiprotons are produced from a quark jet and no
protons are produced from an antiquark jet as in Ref.\cite{adams1}.

Results for $p/\pi^+$ and ${\bar p}/\pi^-$ ratios in central Au+Au
collisions at $\sqrt{s_{NN}}=200$ GeV are shown in the left and
right panels Fig.\ref{ratio}, respectively. Dash-dotted lines are
for the case without conversions between quark and gluon jets, and
they are significantly smaller than those in p+p collisions shown by
dash-dot-dotted lines. Including conversions between quark and gluon
jets increases $p/\pi^+$ and ${\bar p}/\pi^-$ ratios, as shown by
dotted lines, but are still below those from p+p collisions.

The $p/\pi^+$ and ${\bar p}/\pi^-$ ratios are increased or the
difference between the quark and gluon $R_{AA}$ is reduced if the
conversion widths shown in Fig.1 are enhanced by a factor $K_C$. As
shown in Figs.~\ref{raa} and \ref{ratio}, with $K_C=4$ (dashed
lines), similar to the enhancement factor $K_E$ for quark and gluon
drag coefficients, and larger, e.g., $K_C=6$ (solid lines), the
quark and gluon nuclear modification factors become closer and final
$p/\pi^+$ and ${\bar p}/\pi^-$ ratios are also more similar to those
in p+p collisions.

We note that quark and gluon jet conversions have a larger effect on
the ${\bar p}/\pi^-$ ratio than the $p/\pi^+$ ratio. This is due to
the smaller number of antiquark jets than quark jets in heavy ion
collisions and the assumption that quark and antiquark jets fragment
only to protons and antiprotons, respectively. Including some
admixture of proton and antiproton production from quark jets as
well as from antiquark jets would reduce the difference in the jet
conversion effect on $p/\pi^+$ and ${\bar p}/\pi^-$ ratios.

As to the effect of using different fragmentation functions, we do
not expect them to change our conclusions. For example, if we use
instead the KKP fragmentation function \cite{kkp}, which describes
the pion spectrum well but underestimates the proton and antiproton
spectra at high transverse momentum by about a factor of three, jet
conversions are still needed to obtain same $p/\pi^+$ and ${\bar
p}/\pi^-$ ratios in both cental Au+Au collisions and p+p collisions
at same energy. Of course, their values are about a factor of three
smaller than measured ones.

\section{discussions and conclusions}

We have studied the effect of both elastic and inelastic two-body
scatterings of quark and gluon jets in a quark-gluon plasma not only
on their energy loss but also on the conversions between them.
Although inelastic two-body scatterings of quark-antiquark
annihilation and gluon-gluon fusion always lead to conversions
between quark and gluon jets, elastic two-body scatterings can be
non-conversion scatterings if the momentum of the quark or gluon jet
remains to be the one with a larger transverse momentum after the
scattering. To mimic the effect of radiative energy loss of quark
and gluon jets in the quark-gluon plasma, we have multiplied the
drag coefficients calculated from two-body scattering by a factor of
four. We have found that two-body conversion scatterings lead to a
small net conversion of the quark jets to the gluon jets, resulting
in slightly larger $p/\pi^+$ and $\bar p/\pi^-$ ratios than in the case without
including conversions between gluon and quark jets but not large
enough to bring these ratios in central Au+Au collisions to
those in p+p collisions at same energy. A large conversion
enhancement factor of more than four, similar to that needed for the
jet drag coefficients to describe the jet energy loss, is needed to
explain the experimental observations. We note that somewhat smaller
enhancement factors would be needed if the time for starting the
interactions of initial jets with the QGP is earlier, such as the
formation time $\sim 1/p_T$ of the jets, instead of 0.5 fm/$c$ used
in present study. For example, with a starting time of $0.05$ fm/c
reduces the enhancement factors by about 20\%.

In our study, we have assumed that the QGP produced at RHIC is in
chemical equilibrium with about twice many quarks and antiquarks
than gluons. If the produced partonic matter is a pure gluon matter,
resulting $p/\pi^+$ and ${\bar p}/\pi^-$ ratios in central Au+Au
collisions turn out to be slightly smaller than those from a
chemically equilibrated QGP as the gluon matter enhances the
conversion of gluon jets to quark jets via the inelastic scattering
$gg\to q\bar q$.  On the other hand, a pure quark and antiquark
matter gives slightly larger $p/\pi^+$ and ${\bar p}/\pi^-$ ratios
than from a chemically equilibrated QGP as the rate for the
inelastic conversion process $q\bar q\to gg$ is enhanced. As for a
chemically equilibrated QGP, none of these two scenarios is able to
increase the $p/\pi^+$ and ${\bar p}/\pi^-$ ratios at high
transverse momentum in central Au+Au collisions to approach those in
p+p collisions at same energy without requiring a large enhancement
factor for the net quark to gluon jet conversion rate.

Our results that the net quark to gluon jet conversion rate in
central heavy ion collisions at RHIC is much larger than that given
by the lowest-order in QCD may not be surprising as the jet drag
coefficients from the lowest-order QCD need to be enhanced by a
similar factor to describe their energy losses in QGP. Also,
previous studies using the multi-phase transport (AMPT) model
\cite{lin1}, that includes only two-body scattering among partons,
have shown that a much larger parton scattering cross section than
that given by the lowest-order QCD is needed to describe many other
experimental observations at RHIC such as the the large elliptic
flows of hadrons made of light quarks \cite{lin2,chen1} or heavy
quarks \cite{zhang} and the two-pion correlation functions
\cite{lin3}. The large enhancement factor over the lowest-order QCD
results can be considered as an effective parameter for taking into
account effects not considered in present study such as higher-order
contributions and multi-body scattering. A preliminary study of jet
conversions via higher-order radiative scatterings with thermal
partons shows that they lead to similar quark and gluon jet
conversion widths as the lowest-order two-body scatterings
considered in present study. Also, using the QCD coupling constant
at finite temperature as extracted from the lattice QCD studies
\cite{kaczmarek}, which is about a factor of two larger than that
from the perturbative QCD, further increases the quark and gluon jet
conversion widths, resulting in a conversion enhancement factor
close to the needed one. Our results thus could be another
indication for the strongly coupled QGP that has been produced in
central heavy ion collisions at RHIC \cite{starphenix}.

\begin{acknowledgments}
We thank Zhangbu Xu and Carl Gagliardi for helpful discussions. This
work was supported in part by the US National Science Foundation
under Grant No. PHY-0457265, and the Welch Foundation under Grant
No. A-1358. The work of B.W.Z. was further supported by the National
Natural Science Foundation of China under Project No. 10405011 .
\end{acknowledgments}

\end{document}